\documentclass[letterpaper, 10 pt, conference]{ieeeconf}  

\IEEEoverridecommandlockouts                              %
\usepackage{times}
\usepackage{times}
\usepackage{multicol}
\usepackage{subfigure}
\usepackage{floatrow}
\usepackage{graphics} %
\usepackage{epsfig} %

\usepackage[bookmarks=true]{hyperref}

\usepackage{multicol}
\usepackage[ruled,longend]{algorithm2e}
\usepackage[bookmarks=true]{hyperref}

\usepackage[utf8]{inputenc}
 
\usepackage{graphics} %
\usepackage{subfigure}
\usepackage{epsfig} %
\usepackage{mathptmx} %
\usepackage{times} %
\usepackage{amsmath} %
\usepackage{amssymb}  %
\usepackage{multirow}
\usepackage{color}
\usepackage{amsfonts}
\usepackage{xcolor}
\usepackage{tikz}
\usepackage{etex}
\usepackage{arydshln}
\usepackage{footnote}
\usepackage{floatrow}
\usepackage{booktabs}
\usepackage{inputenc}
\usepackage{times}
\usepackage{multicol}
\usepackage{subfigure}
\usepackage{floatrow}
\usepackage{graphics} %
\usepackage{epsfig} %

\usepackage[bookmarks=true]{hyperref}
\usepackage{multicol}
\usepackage[ruled,longend]{algorithm2e}
\usepackage[bookmarks=true]{hyperref}

\usepackage[utf8]{inputenc}
 
\usepackage{graphics} %
\usepackage{subfigure}
\usepackage{epsfig} %
\usepackage{mathptmx} %
\usepackage{times} %
\usepackage{amsmath} %
\usepackage{amssymb}  %
\usepackage{multirow}
\usepackage{color}
\usepackage{amsfonts}
\usepackage{xcolor}
\usepackage{tikz}
\usepackage{etex}
\usepackage{arydshln}
\usepackage{footnote}
\usepackage{floatrow}
\usepackage{booktabs}
\usepackage{inputenc}
\usepackage{txfonts}
\usepackage{hyperref}	

\usepackage{setspace}

\usepackage[draft]{changes}
\definechangesauthor[name={Hemanth}, color=purple]{HM}
\definechangesauthor[name={Amir}, color=blue]{AM}
\definechangesauthor[name={Dantu}, color=red]{KD}
\definechangesauthor[name={Prajit}, color=green]{PR}
\title{\LARGE \bf
Learning Robot Swarm Tactics over Complex Adversarial Environments
}
\author{Amir Behjat$^{1}$, Hemanth Manjunatha$^{1}$, Prajit KrisshnaKumar$^{1}$, Apurv Jani$^{1}$, Leighton Collins$^{1}$,\\ Payam Ghassemi$^{1}$,  Joseph Distefano$^{1}$, David Doermann$^{2}$, Karthik Dantu$^{2}$, Ehsan Esfahani$^{1}$, \\  Souma Chowdhury$^{1}$ $^{\dagger}$%
\thanks{$^{1}$ Department of Mechanical and Aerospace Engineering, University at Buffalo, Buffalo,
NY, 14260 USA.}%
\thanks{$^{2}$Department of Computer Science and Engineering, University at Buffalo, Buffalo,
NY, 14260 USA.}%
\thanks{$^\dagger$ Corresponding Author, 
{\tt\small soumacho@buffalo.edu}}
 \thanks{
 *This work was supported by the DARPA award HR00111920030 and NSF award IIS-1927462. Any opinions, findings, conclusions, or recommendations expressed in this paper are those of the authors and do not necessarily reflect the views of the DARPA or the NSF
}
\thanks {\copyright\space2021 IEEE. Personal use of this material is permitted. Permission from IEEE must be obtained for all other uses, in any current or future media, including reprinting/republishing this material for advertising or promotional purposes, creating new collective works, for resale or redistribution to servers or lists, or reuse of any copyrighted component of this work in other works.}
}

\begin{document}
\maketitle


\begin{abstract}
To accomplish complex swarm robotic missions in the real world, one needs to plan and execute a combination of single robot behaviors, group primitives such as task allocation, path planning, and formation control, and mission-specific objectives such as target search and group coverage. Most such missions are designed manually by teams of robotics experts. Recent work in automated approaches to learning swarm behavior has been limited to individual primitives with sparse work on learning complete missions. This paper presents a systematic approach to learn tactical mission-specific policies that compose primitives in a swarm to accomplish the mission efficiently using neural networks with special input and output encoding. To learn swarm tactics in an adversarial environment, we employ a combination of 1) map-to-graph abstraction, 2) input/output encoding via Pareto filtering of points of interest and clustering of robots, and 3) learning via neuroevolution and policy gradient approaches. We illustrate this combination as critical to providing tractable learning, especially given the computational cost of simulating swarm missions of this scale and complexity. Successful mission completion outcomes are demonstrated with up to 60 robots. In addition, a close match in the performance statistics in training and testing scenarios shows the potential generalizability of the proposed framework.%
\end{abstract}
\begin{figure*}
\centering
\vspace{1.5mm}
\includegraphics[width=0.98\textwidth]{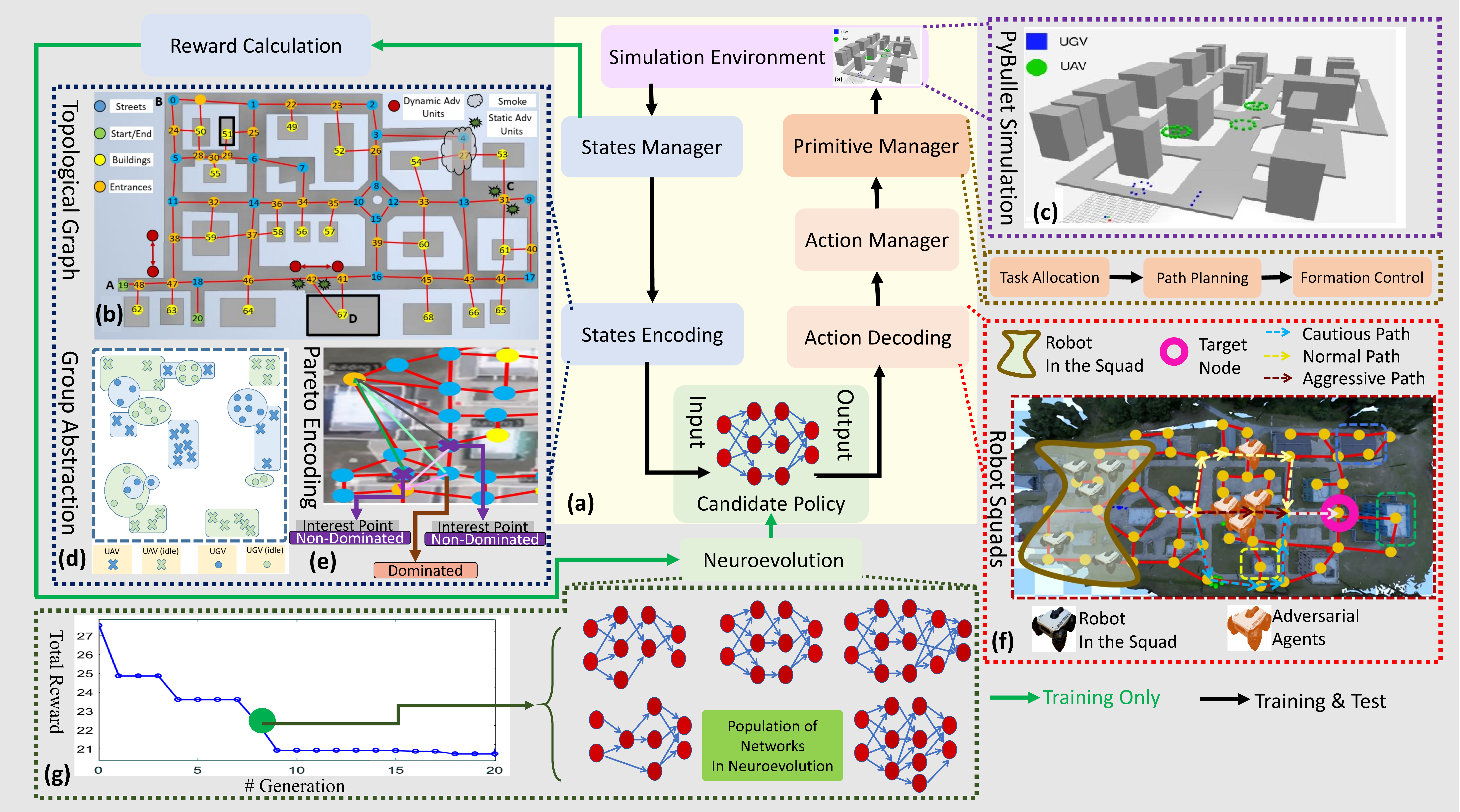}\vspace{-0.1cm}
\caption{Flowchart of the working of our framework. Our framework uses a neuroevolution algorithm called AGENT (a). The whole application is simulated in a custom simulation environment built on top of PyBullet (c). The map is abstracted into a topological graph (b). The swarm is divided into smaller groups that are represented in the group abstraction (d). Using Pareto Optimality, we narrow the points of interest from the topological graph (e). Actions are simplified into applying primitives such as task allocation, path planning, and formation control on groups of robots (f). The measured rewards are fed back into the Neuroevolution module for learning (g).}
\label{fig:overal_frame_work_new}
\end{figure*}

\section{Introduction}

Swarm robotic systems promise new operational capabilities in uncertain and often adversarial environments such as in disaster response, surveillance, and tactical assistance on battlefields~\cite{matsuno2004rescue,AndrewIlachinski2017}. 
Swarm systems require basic capabilities such as dynamically partitioning the swarm into groups and commanding the groups to carry out different tasks that collectively ensure the success of the mission. Adversarial environments present additional challenges to regular swarm applications because the swarms have to make tactical decisions such as opting for suboptimal choices that depend on the decision to ensure safety or sacrifice part of the swarm for the success of the mission. 
Existing approaches for guiding swarm-level decision-making (with 10s-100s of robots) can be broadly divided into handcrafted and automated approaches, with most implementations tested on simplified environments and often over rudimentary tasks as opposed to operationally relevant missions. Handcrafted approaches are algorithms for swarm deployment created by an expert and require expert guidance for every new deployment/scenario. These lack the ability to address various environmental/mission complexities and do not scale well with respect to the size of the swarm due to the need for tedious heuristics for every scenario \cite{AndrewIlachinski2017,brambilla2013swarm}. Automated approaches, such as the use of reinforcement learning (RL) and/or evolutionary computing to design swarm behavior, have mostly been limited to the design of individual swarm primitives such as formation control~\cite{huttenrauch2019deep}, coverage~\cite{jiang2020multi} and target tracking~\cite{baldazo2019decentralized}.  These automated approaches have also been restricted to small, homogeneous teams of robots and relatively simple environments \cite{tan2013research, kim2016design,fan2017evolving,gajurel2018neuroevolution} and it is challenging to scale them up directly. 
Recently, deep RL approaches have also been applied to swarm systems \cite{huttenrauch2019deep} and multi-vehicle problems \cite{stodola2016tactical,mukadam2017tactical,hoel2020tactical,kokkinogenis2019tactical}. Nonetheless, it remains challenging to tackle the large state spaces presented by large-scale swarms operating in complex environments involving obstacles and uncertain adversarial factors. Large input/output spaces, which demand large neural networks-based policy models, are known to cause problems for RL approaches in general \cite{hinton2006reducing, lillicrap2015continuous,gueant2019deep}.

However, ubiquitous realization of swarm robotics in operationally-relevant scenarios needs \textit{tactical intelligence} that can bridge the gap between high-level mission objectives and swarm primitives -- a capability that the current learning or evolutionary approaches rarely provision. We believe that our work is one of the first to provide a methodology to learn such tactic-level policies for robot swarms. We primarily use a recently proposed neuroevolution algorithm~\cite{behjat2019adaptive} for this purpose. Our framework is also capable of incorporating other evolutionary and policy gradient methods, which is demonstrated by comparing the neuroevolution results with a that of a standard actor-critic (A2C) RL algorithm~\cite{grondman2012survey}).
We are able to achieve this by innovating in several aspects: 



\underline{\it Group Abstraction}: A major challenge in observing the state of the swarm for planning is an explosion of the state space. Motivated by grouping explosion strategy~\cite{zheng2013group}, we design our learning system to command groups instead of individual robots and thereby reduce the input complexity. \\
\underline{\it Topological Graph}: Similarly, continuous input/output spaces~\cite{smart2000practical} in swarm operations result in misleading, sparse or delayed rewards~\cite{singh2009rewards} and costs that burden the number of learning samples~\cite{dulac2019challenges}. We address this problem through a topological graph abstraction of the mission environment over which swarm commands are given, which provides a much simpler representation of the state/action spaces.  \\
\underline{\it Pareto Optimality}: Further, we use Pareto optimality to creatively identify a smaller subset of critical points of interest in the action space for improved mission outcomes.\\
\underline{\it End-of-Mission Rewards}: Given the uncertainty of each deployment and the presence of adversaries, it is challenging to design consistent intermediate rewards during the mission. Therefore, we formulate end-of-mission rewards that closely align with mission objectives. \\
\underline{\it Novel Simulation Environment}
\footnote{Public release note: On publication of this paper, we hope to open-source the simulation environment and inbuilt primitives for use by other research groups interested in learning swarm behaviors at scale as well as accurate comparison of current and future methods on a common platform.}: To tackle the end-of-mission nature of rewards, we needed a fast simulation environment that is also reasonably representative of the environment complexity to be used in learning. To this end, we custom-built an environment that uses PyBullet for realistic physics and incorporates several open-source libraries and custom implementations of swarm primitives such as path planning, formation control, and adversary avoidance for group control. We provide an easy way to create and import topological maps for the output space abstraction in this simulator. Most importantly, our simulator runs much faster than real-time for quick learning, even at the scale of 10s of robots and an area of 100s of sq. m.  

To demonstrate these contributions, we consider a swarm mission that deploys 6-40 unmanned ground vehicles (UGVs) and 10-40 unmanned aerial vehicles (UAVs) of two types. This swarm searches a multi-block urban area to find the building (among multiple candidates) that contains a target, while tackling multiple adversarial squads that can engage and disable the swarm robots. While this swarm application was specific to our project, we believe that most of its elements generalize to interesting problems such as search-\&-rescue, disaster response and pollution-cleanup. 


For the rest of the paper, we distinguish between {\it primitives} and {\it tactics} in swarm behavior design. We refer to the single behaviors of a group as a primitive. Examples could be formation control ~\cite{cheah2009region}, distributed mapping ~\cite{fox2006distributed}, signal source localization \cite{ghassemi2020extended}, group coverage \cite{collins2021scalable}, and others. Tactics are ensembles of primitives commanded on multiple groups for swarm behavior. In this work, we assume that group primitives have been previously implemented and are available for use. Thus, the objective of this work is to learn swarm tactics given the primitives for efficient swarm operation in the presence of uncertainty and adversaries.


\label{sc:Introduction}


\section{Swarm Tactics Learning Framework}

\label{sec:Framework}
\label{ssec:framework}


{Figure \ref{fig:overal_frame_work_new} shows the overall framework used for learning. The topological graph, group abstraction, and the Pareto encoding of the graph form the state abstraction and encoding processes producing the effective input features for the neural network-based tactics model. The groups (or squads) and primitives commanded to them form the action decoding, which is the neural network's output. The primitives are managed in the simulation through the primitives' manager. Similarly, the states are obtained from the simulation environment through the State manager. On execution of the scenario, the rewards are calculated and fed back to the neuroevolution algorithm, which evolves a population of neural network models termed as candidate tactics (policy) models in Fig. \ref{fig:overal_frame_work_new}). Iteratively, this framework converges to an efficient neural network that is suitable for the task at hand. 
}

We will now describe each component in more detail. 


\subsection{Encoding State and Action Spaces for Learning}
\label{sec:Encoding}

A major challenge in learning swarm behaviors is state explosion. Realistically modeling the swarm's perception, control, and coordination, and learning over that state space is usually intractable. One of the major contributions of this work is the abstraction of this state-space to make the learning problem tractable while allowing the framework to learn the dimensions of importance to execute the end application efficiently. We use various approaches to reduce the state and action spaces while mitigating loss of information or representation flexibility. These include i) a topological graph abstraction of the urban environment map, ii) Pareto encoding of critical nodes of the graph, iii) modeling the impact of adversarial units, and iv) spatial clustering of the robots to encode the input space. These methods then feed into the input and output encoding processes, which are all further described below.

\subsubsection{Group Abstraction}
\label{ssec:Group}
We define groups of robots in the swarm and command these groups with various primitives. These groups can change over time depending on the mission requirement. Spatial proximity drives the grouping for efficiency. Such grouping is used to construct the state input that is fed into the tactics (policy) model embodied by a feed-forward (tanh) neural network -- {Figure \ref{fig:overal_frame_work_new} \textbf{(d)}}. In Figure~\ref{fig:overal_frame_work_new}, {\it squads} refer to the groups of robots defined by the output or action space of the tactics model. Action determines the relative sizes of squads, Pareto points to be visited by each squad, and the relative (normalized) degree of caution used in (avoidance) path planning in response to adversarial units. By encoding these outputs in relative space, we can generalize the tactics network across mission scenarios and possibly (with some additional tuning) transfer it between urban environments. The actions of the tactics network are processed by the task abstraction layer to form squads of exact sizes (by the dynamic regrouping of the swarm), allocate them to exact destination nodes (in the topological graph) to be visited and provide the de-normalized level of caution to use in avoiding adversarial units.

\subsubsection{Map Abstraction}
\label{ssec:map_graph}
Here we use a graph representation of the map, as illustrated in Figure \ref{fig:overal_frame_work_new} \textbf{(b)}. This graph is generated by encoding locations of interest -- such as intersections, buildings, and building entrances -- to nodes connected via edges with lengths equal to the Euclidean distance between the corresponding data points on the map. By converting the continuous map space into a graph, we can associate the state of a robot (or a cluster of robots) to graph nodes and assign robots (or squads of robots) to visit any node as an action. In other words, this conversion significantly reduces both the input and output space dimensionality. Moreover, it allows focusing on items in the map/environment that are contextually important (e.g., buildings, intersections, etc.). %

\subsubsection{The Pareto Encoding of Nodes}
\label{ssec:Pareto}

There remains an opportunity for further encoding because all nodes do not have equal importance or may not serve as suitable destination points for the robot squads allocated from the tactics model. To this end, we bring in the concept of Pareto optimality, where the critical points for robots to visit are skimmed out using a non-dominated sorting process \cite{srinivas1994muiltiobjective}. {The process is illustrated in Figure~\ref{fig:overal_frame_work_new} \textbf{(e)}.} This process considers the effort required to reach each potential goal and the likelihood of this goal containing the target objective (e.g., victims), assuming that there are always multiple potential target buildings at the start (based on prior intelligence available to the swarm). Only a single building is assumed to actually contain victims, with other potential target buildings being empty. The Pareto filtering process that identifies candidate destination nodes can be expressed as:
\begin{equation}
\label{eq:obj_multi}
k^* = arg\min_{k} \ f_i (k)  =  P(G_l) \times t (X_k \rightarrow G_l), ~~ l=1,2,\ldots N_l
\end{equation} 
Here, $p_l$ is the number of potential targets; $X_k$ represents the spatial location of the $k$-th graph node that can be allocated as a destination to a squad; $t (X_k \rightarrow G_i)$ is the time taken to reach a potential target $G_i$ from the point $X_k$, and $P(G_l)$ is the probability that the target $G_l$ contains the target objective (victims). The probability is calculated based on the search progress of the UAVs and UGVs, with the initial probabilities being all equal to $1/N_l$. Once the correct target (say $G_l^*$) has been identified (through simulated indoor search), $P(G_l^*)=1$ and $P(G_{l})=0,~\forall l\ne l^*$. {Here, indoor search is the search conducted by UGVs and outdoor search is conducted by UAVs.} Conversely, if one potential target building (say $G_l$) has been (indoor) searched and victims are not found there, then we set $P(G_l)=0$ and $P(G_{l})=1/(N_l-1),~\forall l\ne l^*$.

\subsubsection{Adversaries}
\label{ssec:adversaries}
Real-world conditions contain several types of hazards that we label adversaries. As an illustration, motivated by post-disaster and combat environments, we consider three types of adversities: 1) smoke, 2) static adversarial unit, and 3) dynamic adversarial squads. We should note that our framework can incorporate other adversaries as long as their properties can be accurately modeled in the simulation environment. We model the impact of smoke by reducing the speed of the robots while passing through it. A static adversarial unit is location-based; it cannot be observed or neutralized by the swarm robots and causes complete failure of a robot that comes within a threshold distance of its location. The dynamic adversarial units, which can also cause the failure of swarm robots, appear as squads with specific paths. These adversarial units are observable when in the perception range of the swarm robots and can be neutralized by the swarm robots based on an engagement model. At the same time, swarm robots can also take circuitous paths to avoid these dynamic adversarial units, depending on the degree of caution set by the tactics model, thus leading to interesting tactical trade-offs. As described later in Section \ref{sec:pybullet}, this avoidance behavior is encapsulated by a new primitive that adapts the path planning process.

\subsubsection{Input Encoding}
\label{ssec:Input_encoding}


There are three major types of inputs: 
(i) state of the robots w.r.t. the environment, (ii) state of our knowledge of the environment w.r.t. target buildings, and (iii) mission state. Mission state is represented in terms of remaining time, assuming that the total allowed time is defined in each mission scenario. The knowledge state is represented by probabilities, $P(G_i),~i=1,2,\ldots,p_l$, of the building to be the actual target building, i.e., the one that contains the target objective. 

To encode the input state of the robots w.r.t. the environment (targets and adversaries), we perform K-means clustering of the robot locations separately for each type of robot. We consider one type of UGV and two types of UAVs in our experiments, with each type of robots allowed to form 3 clusters, leading to 9 clusters in total. Figure \ref{fig:overal_frame_work_new} illustrates the clustering of the UAVs and UGVs using the position information extracted from the environment in the state encoding module. Post clustering, the input state attributed to robots is encoded as i) size of each cluster, ii) distance of each cluster's centroid to each Pareto node ($k^*$), iii) distances of clusters to each other, and iv) normal distance of each dynamic adversary to the straight lines connecting clusters to potential target buildings. The last item here seeks to encode the impact of the adversaries on likely future actions based on the current state of the swarm.
Table \ref{tab:inp_outp_list} lists the inputs to the tactics model and their size.

\begin{table}[b]
\begin{center}
\caption{{The State and Action parameters of the tactics model for a swarm with six UAV Squads and three UGV Squads}}\vspace{-4pt}
\setlength\tabcolsep{2pt}
\footnotesize
\label{tab:inp_outp_list}
\begin{tabular}{|p{1cm}| p{5cm} | c|}
\hline
\textbf{} &\textbf{Parameter} & \textbf{Size}  \\
\hline
\multirow{1}{*}{\textbf{Input }}  & {1) Remaining time} & 1    \\ 
\multirow{1}{*}{\textbf{States}} &{2) Prob. of the Target to be the Actual Target} & 3 \\ 
 & {3) Size of UxV Clusters} &  $3 \times 3 $ \\  
   & {{4) Distance of UxV Clusters to each other}} & $  3 \times 3$    \\ 
 & {5) Dist. of UxV Clusters to Pareto Nodes} & $ 3 \times 3 \times (5 + 3)$    \\  
 & {{6) Normal Dist. of Adversaries to Lines Connecting UxV Clusters to Target Bldgs}} & $ 9 \times 3 \times 2$ \\

 \hdashline
 & \textbf{Total \# of Inputs} & {\textbf{148}} \\
 \hline
\multirow{1}{*}{\textbf{Output}} &  {Node to visit for each UxV Squads} &  $3 \times 3 $   \\
\multirow{1}{*}{\textbf{Actions}}   & {Size of each UAV/UGV Squads} &   $3 \times 3 $ \\
    & {Degree of Avoidance of Dyn. Adversaries} &   $3 \times 3 $  \\
  \hdashline
  & \textbf{Total \# of Outputs} & {\textbf{27}} \\
\hline
\end{tabular}
\end{center}
\end{table}

\subsubsection{Output Encoding}
\label{ssec:Output_encoding}
The squads ($m$) are kept fixed for computational tractability across different swarm sizes. Here the output of the tactics model is encoded into three types of actions: \textbf{1)} the number of robots, $N$, in each squad ($s_i,~i=1,2,\ldots m$), \textbf{2)} the node to be visited by each squad ($k_{s_i}$), and \textbf{3)} the degree of caution (avoidance) to be used by each squad w.r.t dynamic adversaries ($\gamma_j$). Output 1 is encoded in relative terms allowing scalability to any swarm size. Output 2 uses the Pareto encoding of the graph space of the mission environment, i.e., only Pareto nodes can be assigned as squad destinations. To generalize the tactics network across different environments and mission scenarios that could lead to different total numbers of Pareto nodes, we fix the size $K$ of the set of nodes assigned as squad destinations. To this end, we apply the crowding distance criteria \cite{deb2002fast} to the Pareto set to further filter out the most diverse set of $K$ nodes that can serve as squad destinations.

\subsection{MDP formulation}\label{ssec:MDP_formulation}
{The problem can be modeled as a Markov decision process (MDP) \cite{garcia2013markov} problem. In the current formulation, the states, actions, and rewards can be defined based on Eq. \ref{eq:MDP}.}

\begin{equation}\label{eq:MDP} 
    \begin{aligned}
     &s = \mathcal{F}_s(N_{C,i},\ \Delta(C_{C,i},X_{k^*}), \ P(G_l),\ t)\\
     &a = \mathcal{F}_a(N_{S,j},\ X_{k^*_j},\ \gamma_j)\\
     &r = \mathcal{F}_r(\Delta(C_{C,i},\ G_l), \psi_l,\ N_{C,i},\ t)\\
    \end{aligned}
\end{equation}
Here $N_{C,i}$ is the size of the $i^{th}$ cluster, $\Delta(C_{C,i},X_{k^*})$ is the distance between the center of that cluster and the Pareto node location $X_{k^*}$, $P(G_l)$ is the probability of each goal $G_l$, and $t$ denotes time elapsed since the start of the mission. Similarly, $N_{s,j}$ is the size of the $j^{th}$ squad, $k^*_{j}$ is the Pareto node to be visited by this squad, and $\gamma_j$ is the \textit{cautious level} with which the squad must proceed. Finally, the reward is defined as a function of four terms: distance between clusters and the goals, size of the remaining clusters, the progress of searching each target ($\psi_l$), and the remaining time.

{The state set comprises different terms itself, including:}    
\begin{equation}\label{eq:MDP_states} 
    \begin{aligned}
     &\mathcal{F}_s = \begin{cases}
s_1 = {(t_{f} - t)\times V_{\max}} / {\sqrt{\Delta_X^2+\Delta_Y^2})},\\
s_{2, \cdots 4}= P(G_l),\\
s_{5, \cdots 13}= {N_{C,i}}/{N_{0}},\\
s_{14, \cdots 22}= |{C_{C,i}},\ {C_{C_j}}|,\\
s_{23, \cdots 84}= |{C_{C,i}},\ X_{k^*}|,\\
s_{85, \cdots 138}= {|(G_l - C_{C,i}) \times (r_o - C_{C,i})|}/{|(G_l - C_{C,i})|},\\
\end{cases}
    \end{aligned}
\end{equation}
{Here $t$ and $t_f$ are the elapsed time and the total allowable mission time; $\Delta_x, \Delta_Y$ are the size of the map, and $V_{\max}$ is the maximum moving speed out of all robots (in this case, UAVs).}
{The first state variable is the remaining time ($t_{f} - t$) normalized by the required time to travel across the map ($\frac{V_{\max}}{\sqrt{\Delta_X^2+\Delta_Y^2}}$). This term allows perception of the required time to allow more or less aggressive actions. 
The second state variable-set is associated with the probability of having a goal $(P(G_l))$ in each target location, $G_l$. The third state variable sub-set gives the relative size of each dynamically abstracted cluster-$i$, with $N_C,i$ being the number of agents in that cluster and $N_0$ being the total number of agents of the respective type at the start of the mission. 
Another feasible choice for normalizing would be the predicted size of the required team (${N_0}^*$).
The fourth and the fifth sub-sets define the distance of robot cluster centers $C_{C, i}$ to each other and the Pareto front points $P_j$ (points of interest). Here we use 5 Pareto front points based on distance (caution level $\gamma = 0$) and 3 other points based on cautious path planning (caution level $\gamma = 1$). The last sub-set finds the normal distance of
\textit{observed adversarial agents}, $r_o$, to the straight line connecting robots in the swarm to the targets.} 

{The action set includes the following different parameters for each squad:}
\begin{equation}\label{eq:MDP_actions} 
    \begin{aligned}
     &\mathcal{F}_a = \begin{cases}
a_{1 \cdots 9} = k^*_j,\\
a_{10 \cdots 18} = {N_{S,j}} / {\sum_{i=1}^{N_C}{N_{C,i}}},\\
a_{19, \cdots 27}= \gamma_j,\\
\end{cases}
    \end{aligned}
\end{equation}
where $K^*_j$ is a generic Pareto node, and $N_{S}$ and $N_C$ are respectively the sizes of squads and clusters.

{An illustration of actions is shown in Fig. \ref{fig:overal_frame_work_new} \textbf{(f)}, where a squad of robots is chosen to go to the desired node and follow a path planning algorithm based on the location of the adversaries.}

{The reward is defined based on rescue time and the casualties with a soft constraint for failure. }

{Due to the stochastic nature of the problem, a Partially Observable Markov Decision Process (POMDP) \cite{spaan2012partially} formulation would have been more appropriate. However, we follow the more straightforward MDP formulation to preserve the tractability of the problems which could have been compromised by the increased computational complexity of the POMDP formulation. }

The choices about the number of clusters ($N_C=3$ for each type of robot), number of squads ($N_S=3$ for each type of robots), and number of Pareto nodes ($K=5+3$) were made to keep the problem size tractable, and were also found to work well for the given number of goals ($N_l=3$) in our case studies. However, in the future, we could make these choices  adaptive to other environmental factors such as the size of the map, the number of possible targets, and human intent regarding how soon the mission must be completed.


\subsection{Learning Algorithm: \textit{Adaptive Genomic Evolution of Neural Network Topologies (AGENT)}:} 
Neuroevolution is a direct policy search method. Unlike policy gradient type RL approaches, neuroevolution uses a specialized genetic algorithm to evolve neural network models that map states to actions for a given problem. We specifically use the AGENT \cite{behjat2019adaptive} algorithm, which falls into the class of neuroevolution methods that simultaneously evolve the topology and weights of neural networks. AGENT builds upon the neuroevolution of augmenting topologies or NEAT paradigm \cite{stanley2002evolving} and its variations \cite{stanley2019designing}. We chose to use AGENT because of its demonstrated adaptive control of reproduction operators and 2-way variation in topological complexity \cite{behjat2019adaptive}, which can help alleviate stagnation issues otherwise affecting earlier neuroevolution methods.


During the neuroevolution process, here, every candidate tactics model is evaluated over a set of mission scenarios, with its fitness given by the mission-level reward function aggregated over these scenarios: $\mathcal{F} = {\sum \limits_{sc=1}^{N_{sc}}  R_{sc}}/{N_{sc}}$.


\section{Swarm Simulation Environment}
\label{sec:pybullet}

While the abstractions simplified the learning problem, we observed that it would still take a very long time to learn swarm tactics if we used simulators with 3D physics, such as ROS or Gazebo. Therefore, we designed a new simulator (that's also used for human-swarm interaction studies \cite{manjunatha-SMC2020}) to perform simulations that run much faster than real-time.  

To build our swarm simulation, we use the open-source PyBullet~\cite{coumans2019} library. The full-scale 3D computer-aided model layout of a $300m \times 150m$ urban area (as shown in Fig.~\ref{fig:overal_frame_work_new}\textbf{(b)}) is developed using Autodesk Inventor and imported into Pybullet through URDF. 
The simulation considers only the kinematic behavior, implemented as a positional constraint between the robot (UAV or UGV) and origin. The simulation has a headless mode without UI for fast execution and a GUI mode for visual analysis and demonstration \cite{manjunatha-SMC2020}.

\subsection{Primitives}
\label{ssec:Primitives}

{We utilized different primitives in the paper. The main primitives included in our experiments are:}

\subsubsection{Task Allocation}
\label{ssec:Primitives-task_alloc}
A simplified task allocation approach is used for computational tractability, executed at each tactical decision-making instance. It divides the swarm into 9 squads (3 squads each for the two types of UAVs and 3 squads of UGVs). The task allocation algorithm then separates the idle robots from non-idle robots and randomly distributes the task among idle robots. 



\subsubsection{Path Planning} \label{ssec:Primitives-path_planning}

\begin{algorithm}
\footnotesize{
  $I$: Segmented image of the environment;
  $G$: Graph\;
  $S \gets$ Skeletonize image I\;
  $G.\texttt{init}(x_\text{init})$\;
  \For{\texttt{pixel} in \texttt{medial-axis of} S}
  {
    $x_\text{new} \gets \text{\texttt{pixel position}}$\;
    $x_\text{edge} \gets Edge(x_\text{init},\ x_\text{new})$\;
    $G.\texttt{addnode}(x_{\text{new}})$,\ $G.\texttt{addedge}(x_{\text{edge}})$\;
    $x_\text{init} \gets x_{\text{new}}$\;
    
  }
  \KwRet{$G$}\;
  }
  \caption{The skeleton graph for path planning}
  \label{algo:path_planning}

\end{algorithm}

\begin{algorithm}[b]
\footnotesize{
  $G.\texttt{init}(x_\text{init})$\;
  \While{$\texttt{Scenario is not done}$}
  {
    \For{$\forall~ \texttt{Smoke (}D_i\texttt{)in Environment}$}
    {
        \If{$D_i\texttt{ is observed}$}
        {
            \For{$\forall e_j \in G | e_j \in D_i$}
            {
                $e_j = e_j \times (1 + f({D_i}, e_j))$\;
            }
        }
    }
    \For{$\forall~ \texttt{Enemy (}L_i\texttt{)in Environment}$}
    {
        \If{$L_i\texttt{ is observed}$}
        {
            \For{$\forall e_j \in G | e_j \in L_i$}
            {
                $e_j = e_j \times (1 + \texttt{Caution Level}, e_j))$\;
            }
        }
    }
  }
  \caption{Updating skeleton graph w/ adversaries}  
  \label{algo:update_path}
  }
\end{algorithm}

{To implement the path planning primitive, we extract the continuous positions on the map in the form of the graph, as shown in Figure \ref{fig:overal_frame_work_new}\textbf{(b)}. To extract the node graph of pathways from the 3D environment, we extract the occupancy grid -- highlighting only the roads -- and skeletonize the image to obtain the medial axis. This medial axis is then converted into a graph structure \cite{xiaolong2019}. Finally, the resulting graph network is used to query the shortest distance between the current position of a given squad ({using Dijkstra's Algorithm \cite{magzhan2013review}} ) and the desired location on the map (Algorithm \ref{algo:path_planning}). To model the impact of and tactical response to adversaries, we update the skeleton graph based on the location of the smoke and dynamic adversarial units that our swarm robots have observed. %
Here, the weight of each edge (inter-node travel cost) is updated to reflect the impact of smoke and dynamic adversarial units; i.e., $w_{ij} = w_{ij} + \Delta w$. For smoke, the $\Delta w$ is computed as a linear decay function of the radial distance from the location of the smoke unit (up to a set radius), scaled by a set constant. %
For dynamic adversarial units, the weight adaption ($\Delta w$) of each edge is directly given by the third output of the tactics model (see Section~\ref{ssec:Output_encoding} and Table~\ref{tab:inp_outp_list}). Algorithm \ref{algo:update_path} explains this procedure. This primitive is also able to update the map based on the observed enemy units. }

\subsubsection{Formation Control}\label{ssec:Primitives-formation}

{A region-based formation control method~\cite{cheah2009region} is used here. It is a decentralized formation control technique that is useful for controlling large swarms of robots. It uses two components weighted by prescribed coefficients, respectively eliminating inter-robot collision and bounding robot motion within the desired region. Using this method, complex shapes can be formed that satisfy given geometric constraints. Algorithm \ref{algo:formation_control} explains this procedure.}

\begin{algorithm}
\footnotesize{
  \For {$\text{each robot  in the swarm}$}
  {
      $X \gets \text{Position vector of the neighbours}$\;
      $d_{min} \gets \text{Minimum distance}$\;
      $S\gets \text{Shape parameters}$\;
      $H_i \gets\text {Component to avoid inter robot collision} $\;
      $F_i \gets\text {Component to maintain robots inside region}$\;
      \For{$\text{neighbour  in X}$}
      {
        $d_{ij} \gets \text{Distance to neighbours} - d_{min}$\;
        $x_{ij} \gets\text {Relative position vector}$\;
        $H_i = \max(0,\ d_{ij}){x_{ij}} + H_i$\;
      }
      \For{$\text{constrain  in Total constraint}$}
      {
        $f_{i} \leq 0 \gets \text{check constraint}$\;
        $F_i = \max(0,f_{ij}){ x_{ij}} + F_i$\;
      }
      $Velocity  = Path\ velocity + \alpha H_i + \beta F_i$\;
      \KwRet{$Velocity$}\;
    }
  \caption{The formation control algorithm}
  \label{algo:formation_control}
  }
\end{algorithm}

\section{Case Study: Urban Search \& Rescue Mission}
\label{sec:Problem_df}
We study an urban search and rescue mission in a combat environment with these components in the backdrop. This environment includes complexities such as smoke and static and dynamic adversarial squads. We assume a set of UAVs and UGVs with the following capabilities: i) UAVs: have a maximum speed of 5 m/s and are capable of identifying potential and true target building; and ii) UGVs: move with a maximum speed of 1 m/s and are capable of identifying potential and true target buildings, and reaching the victims inside the building (for rescue), which is a requirement for successful mission completion. Furthermore, both UAVs and UGVs are capable of engaging/neutralizing dynamic adversaries, with UGVs possessing a greater probability of neutralizing. Finally, we assume an initial map of the environment is known apriori, while the locations of victims and adversarial units are unknown.%

\noindent\textbf{Mission Objective $\to$ Reward Function Formulation:} The swarm tactics model is trained by maximizing the following objective (or reward) function, which measures the performance of the policy over $N_{sc}$ different scenarios.
\begin{align}\label{eq:obj_eq}
\max_{\theta_\text{NN}} f &= \sum \limits_{sc=1}^{N_{sc}}\left[ \delta_{sc}\times\tau_{sc} \times (\Lambda_{sc})^{C_S} \ + (1-\delta_{sc})\times \sum\limits_{l=1}^{N_{l}}\Psi_{sc,l}\right]
\end{align}
where $\delta_{sc}$ is set at 1 if scenario-$sc$ is successful, i.e., robots rescue the victims in a given maximum allowed mission time. Otherwise, $\delta_{sc}$ is set at 0 (if the scenario-${sc}$ failed). In Eq.~\eqref{eq:obj_eq}, timeout (mission failure) is added to the mission success metric as a soft constraint. { The parameters of the neural network (policy model) are specified as $\theta_\text{NN}$, which includes its structure, weights, and biases.}

If scenario-${sc}$ is successful, the tactics policy is incentivized to finish the mission faster with lower casualty of the swarm robots (higher survival rate). The survivability coefficient $C_S\in[0, \infty)$ balances the survival rate and the rescue time -- a higher value of $C_S$ puts greater importance on the survival rate. The rescue time ($\tau_{sc}$) is the time taken to rescue the victims, normalized by maximum allowed mission time. The survival rate, $\Lambda_{sc}$, is the ratio of the number of surviving robots (end of mission) to the swarm size at the start of the mission in scenario-${sc}$. 
If scenario-${sc}$ fails, the tactics policy is rewarded based on the search progress made. The search progress ($\Psi$) is computed as follows: 
\begin{align}
\Psi_l &= (1 - N_{sc})  +  \frac{1}{N_l}\sum \limits_{l=1}^{N_l}  \frac{\psi_{l, \texttt{in}} +  \psi_{l, \texttt{out}}-2}{2}
\end{align}
Here, the terms, $\psi_{l, \texttt{in}}$ and $\psi_{l, \texttt{out}}$ are indoor and outdoor search progress over $N_G$ potential target buildings. {The search progress  $\psi_{l, \texttt{in}}$ can be estimated by the area searched by robots compared to the whole search area for each building either inside or outside of the building.} The term $(1-N_{sc})$ prioritizes mission success over rescue time.%


\section{Results and Discussion}
\label{sec:results}
We study our learning framework by simulating three distinct experiments, listed in Table~\ref{tab:scenarios}, and defined as: \textbf{\textit{Experiment 1:}} {study the network topology variation, learning convergence, and generalizability for scenarios with no dynamic adversaries, to showcase the overall viability of the neuroevolution-driven swarm tactics generation framework}. \textbf{\textit{Experiment 2:}} {evaluate the trade-offs between rescue time and survival rate, over training and testing scenarios (incl. prolonged operations that use the surviving swarm for sequential search \& rescue missions}. \textbf{\textit{Experiment 3:}} {demonstrate the utility of input and output encodings and their impact on scalability with swarm size, both with independent and collective ablation tests. For all the test cases, we have used the sample scenario as shown in Figure \ref{fig:overal_frame_work_new}\textbf{(c)}. }
All simulations are run by executing a parallel version of our learning framework on a workstation with 2 Intel Xeon Gold 6148 processors (each w/ 20 core CPUs) and 192 GB RAM.
In creating the pool of experiment scenarios, environmental parameters such as the location of target buildings and adversarial agents are designed manually to allow a sufficient environment complexity to be encountered by the swarm. However, note that the training and testing scenarios are randomly chosen from this designed pool of scenarios.

\subsection{Experiment 1: Learning Curve}
\begin{figure}
\vspace{2mm}
\centering
\includegraphics[keepaspectratio=true, width=.95\textwidth]{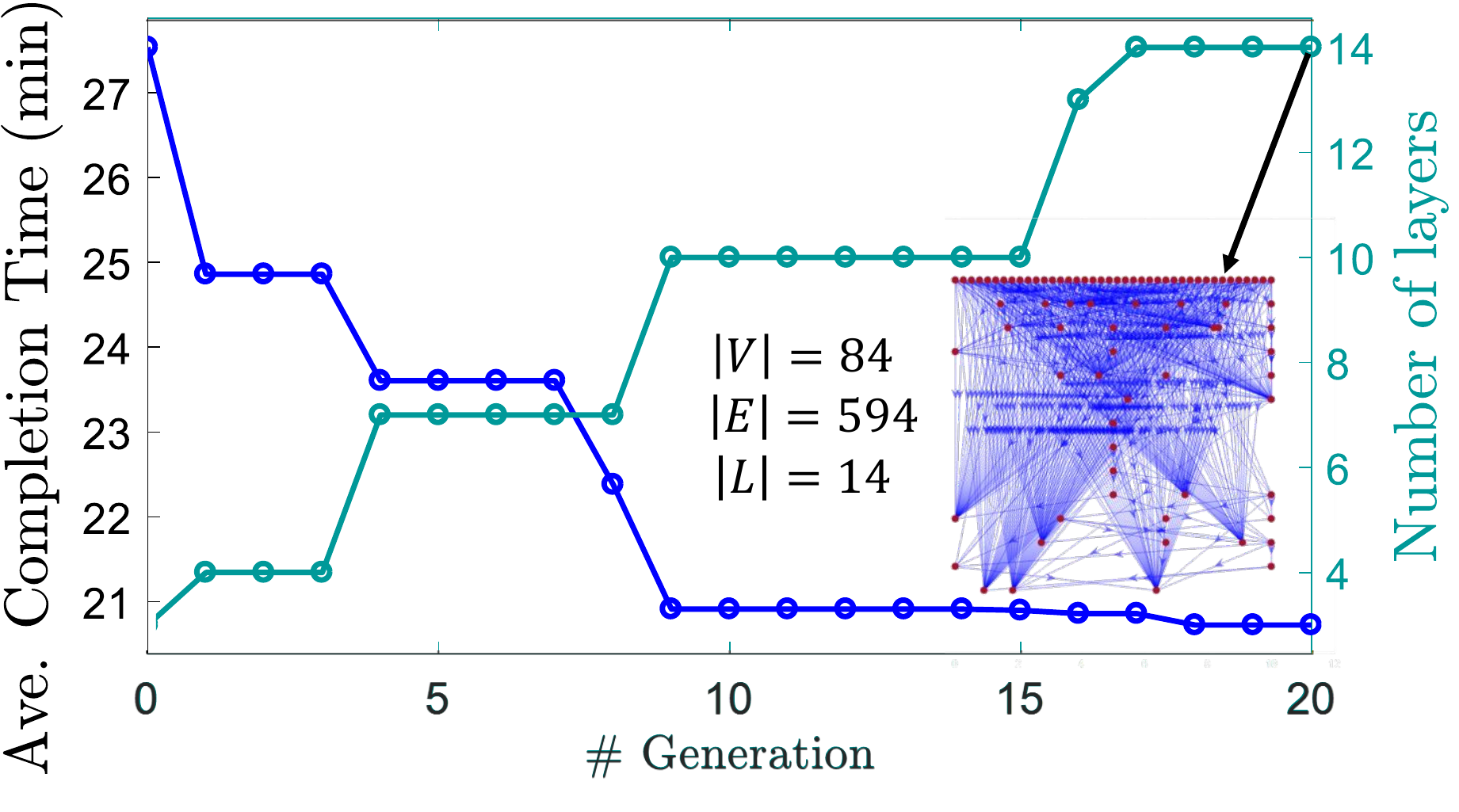}\vspace{-0.2cm}
\caption{Experiment 1: The convergence history of learning with AGENT}
\label{fig:convergence_older}
\end{figure}

In the first experiment, we assess the learning performance using a population of 36 genomes, each evaluated over 15 (training) mission scenarios, with neuroevolution allowed to run for a maximum of 20 generations. These settings  increased the diversity of scenarios while keeping the problem computationally tractable. The training scenarios consist of 3 squads of UAVs and 3 squads of UGVs, and dynamic adversarial units are not included in this case. Subsequently, $C_S$ is set at zero to focus purely on rescue time. The convergence history in this case, as illustrated by Fig. \ref{fig:convergence_older}, provides a promising reduction of 7 minutes of mission completion time (~25\% improvement) compared to the initial randomized policies. In this process, the network complexity increases from roughly 3 to 14 layers (estimated), with the final network (comprising 84 nodes and 594 edges) shown as a docked diagram inside Fig. \ref{fig:convergence_older}.  %

Concerning the rescue time in minutes, the trained tactics model yields a mean of 20.72 and a standard deviation of 1.23. Furthermore, across the 6 unseen test cases, we observe a mean rescue time (in mins) of 20.50 and a standard deviation of 1.01, thereby demonstrating the potential generalizability of the learning framework. 

\subsection{Experiment 2: Analysis of the Survivability Coefficient}
This section studies the impact of the survivability coefficient on the learned tactics model over two different test conditions. Here, we run the learning for two different values of $C_S$ in the reward function, as follows: 1) $C_S=0$: the reward function is only a function of the rescue time, and 2) $C_S=1$: the reward function is a function of both rescue time and survival rate. Table~\ref{tab:scenarios} summarizes the range of values used to generate training and test scenarios. A set of 15 and 54 different scenarios are generated, respectively, as training and test sets by combining different numbers of robots and adversarial units (incl. smoke, static and dynamic adversarial units) and different distribution of potential target buildings. The map of the environment is kept the same for all training and test scenarios. The maximum allowed mission time is set at 40 minutes. The learning algorithm is executed for 10 generations with a population of 36 genomes.

\begin{table}
    \begin{center}
    \caption{Scenario Description}
    \footnotesize
    \label{tab:scenarios}
    \setlength\tabcolsep{1.5pt}
    \begin{tabular}{|c| c | c| c | c | c |}
    \hline
    \textbf{Exp}  & \textbf{\# Scenario}  & \textbf{\# UAV, \# UGV} & \textbf{\# Adversaries} & \textbf{Smoke} \\
    \textbf{Study} & \textbf{Train, Test} &  & \textbf{Static, Dynamic} & \textbf{Radius} \\
    \hline
    \textbf{1} & 15, 0 & $10-40,\ 10-40$ & $2,\ 0$ & 10m \\
    \hline
    \textbf{2, 3} & 15, 54 & $12-36,\ 6-24$ & $0-6,\ 0-14$ & 0-10m \\ 
     \hline
    \end{tabular}
    \end{center}
    \vspace{-6pt}
\end{table}

For each learned tactics model, we evaluate its performance on the 54 unseen test scenarios with two different test conditions: \textbf{1) Test with Independence (Test-I):} Each scenario is independent of other scenarios (each scenario is treated as one mission with termination), and \textbf{2) Test with Continuation (Test-C):} Each scenario is executed using surviving units from the previous scenario (continuation). Fort Test-I, the number of UAVs and UGVs are in a range of $[12,36]$ and $[6,24]$, respectively. In the \textit{test with continuation}, the number of UAVs and UGVs are set at the maximum value (i.e., 36 UAVs and 24 UGVs) at the start. As the simulation continues, the number of agents available in each scenario is the same or less than that in the previous scenario because of losses/casualties due to encounters with adversaries.  

\begin{figure} %
\centering
  \subfigure[Average mission Rescue time]{\includegraphics[width=0.47\textwidth]{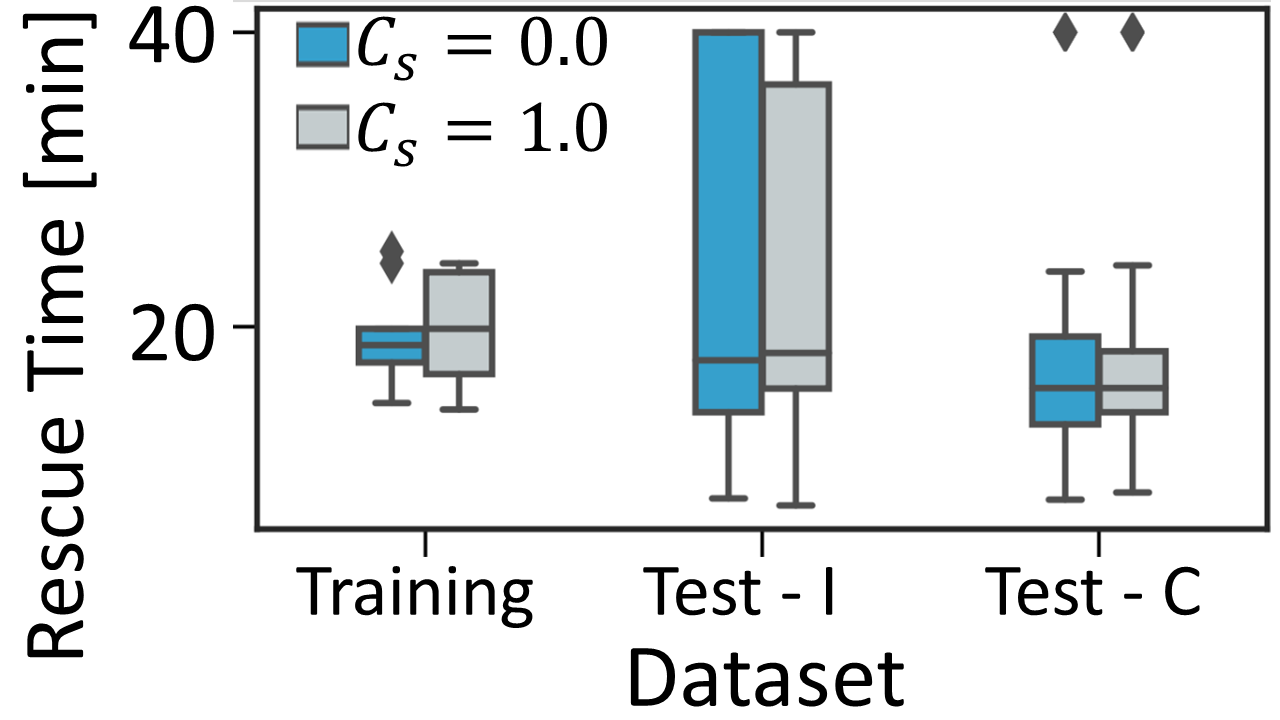}\label{fig:times_box}}~
  \subfigure[Average mission Survival rate]{\includegraphics[width=0.47\textwidth]{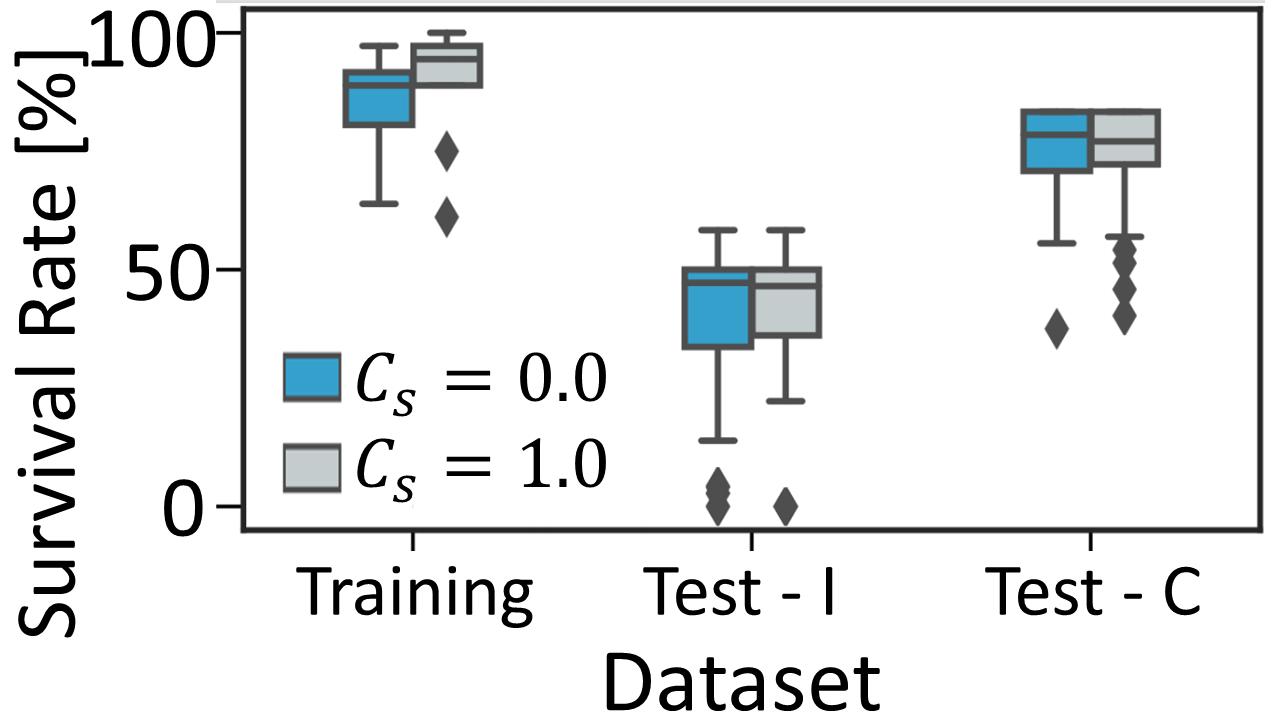}\label{fig:survival_box}}
\vspace{-3mm}
\caption{Experiment 2: The mission performance for three sets of scenarios: Training, Test, and Continuing Test Scenarios w/o replacement of robots.}
\label{fig:performanceAnalysis}
\vspace{-5pt}
\end{figure}

The performance of each learned model ($C_S=0$ and $C_S=1$) in terms of both mission rescue time and survival rate are shown in Figs.~\ref{fig:times_box} and \ref{fig:survival_box} respectively. It can be seen that the rescue time and the survival rate increase on both test conditions by changing the survivability coefficient ($C_S$) from 0 to 1. The improvement in survival rate is expected, but the increase in both metrics can be explained as follows: with the survivability coefficient set at 1 ($C_S=1$), the tactics model explores to find less dangerous solutions, and if the shorter paths in the training set include adversaries, then the tactics model prefers larger path deviation and limits encounters. %

Figures~\ref{fig:times_box} and \ref{fig:survival_box} show a significant difference in the performance of the learned tactics, in terms of both rescue time and survival rate, over two different test conditions. We observe a higher survival rate in Test-C in comparison with Test-I. One explanation for this observation is that the effectiveness of the engagement of dynamic adversarial teams is modeled such that it decreases with an increasing number of robots that they are engaged with. In Test-C, a larger number of robots are available, which can alleviate the effect of adversaries, and the survival rate is higher. In addition, due to lower casualties in Test-C, the robots can identify the target and rescue the victims sooner. On further analysis of the tactical level behavior, we observe a positive correlation between the average adversarial unit avoidance of robots and mission rescue time and survival rate.

\subsection{Experiment 3: Ablation Study on Encoding}
We perform an ablation study to analyze the effect of input and output encoding by creating the following four cases: \textbf{1) Both (input \& output):} In this case, we ablate both the input and output encoding. \textbf{2) Input:} The clustering procedure in the input encoding is removed, and the states of robots are directly fed into the tactics model; However, we keep the output encoding. \textbf{3) Output:} The Pareto filtering approach is removed from the output encoding while keeping only input encoding. \textbf{4) None:} This represents our primary framework with both input and output encoding. For brevity, we only report the results of the reward function with $C_S = 0$ (i.e., the survival rate is omitted). The settings are listed in Table~\ref{tab:scenarios}. 

\begin{figure}
\centering
\vspace{2mm}
    \includegraphics[width=0.95\textwidth]{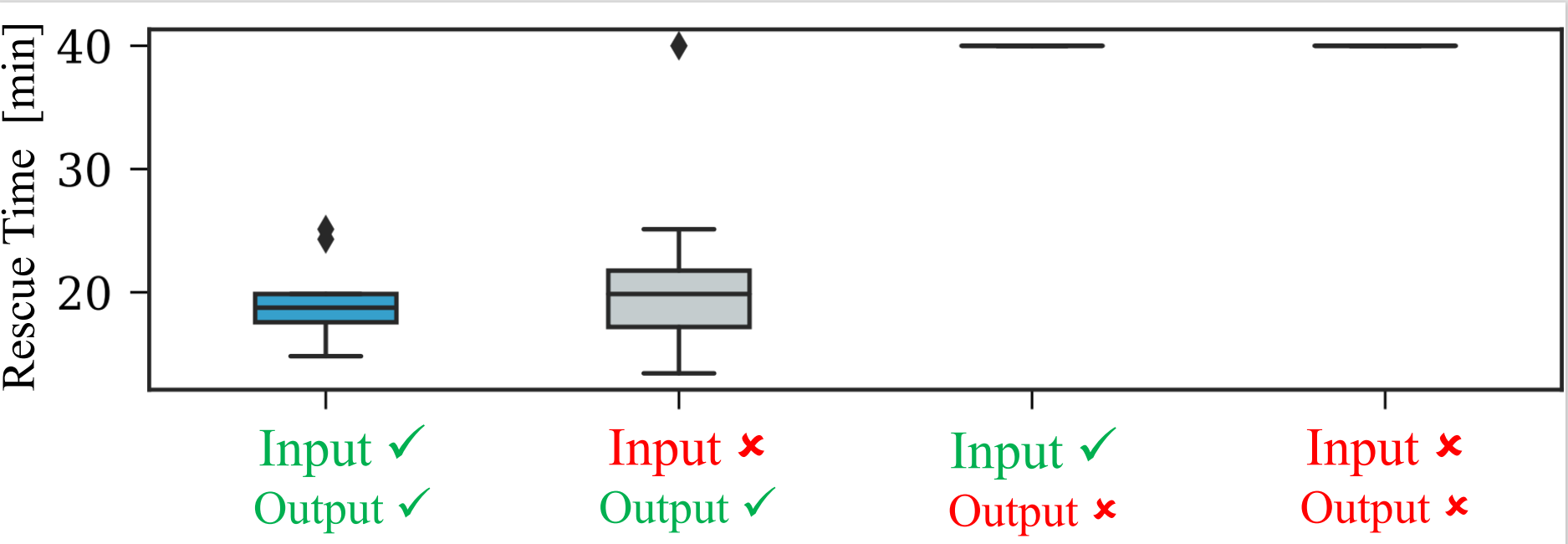}
\caption{The effect of input/output encoding on swarm tactics learning using Neuroevolution for four encoding conditions.}
\label{fig:box_no_dncoding_performanceAnalysis}
\end{figure}

Figure \ref{fig:box_no_dncoding_performanceAnalysis} summarizes the results of each case in terms of rescue time over unseen test scenarios. Our results illustrate the remarkable importance of output encoding (the Pareto filtering approach). When learned without output encoding, the resulting tactics model cannot complete any mission within the max allowed mission time (40 mins). This observation is expected since without an output encoding the robot action space will be quite large. On the other hand, input encoding is not as critical for learning successful tactics. Having said that, the input encoding is still important. First, the input encoding improves the quality of the learned tactics. For example, the full encoding shows 10.5\% improvement in average performance and a 60\% improvement in variance compared to the output-only encoding. Secondly, and most importantly, the input encoding enables the generation of tactics relatively invariant to swarm size. 
\vspace{-4mm}

\subsection{Reinforcement Learning}
\label{ssec:RL}

To demonstrate the generality of our framework, we replace Neuroevolution with a popular gradient-based reinforcement learning (RL) method, the actor-critic algorithm or  A2C~\cite{grondman2012survey}, to learn suitable tactical policies. 
Table \ref{tab:A2C_params} lists the parameters used for implementing RL. Due to the computational burden  of running A2C, we limited our simulations to three training scenarios and 15 testing scenarios. All the scenarios are from the same pool as those previously used for neuroevolution. With RL, learning is again executed with and without encoding. A total of 300,000 time steps of training are allowed for A2C. The policy network is defined to be a multi-layer perceptron with two  hidden layers with 64 neurons per layer and an output action layer.

\begin{table}
\begin{center}
\caption{A2C Parameters used in the Ablation Study of Encoding}\vspace{-4pt}
\footnotesize
\label{tab:A2C_params}
\setlength\tabcolsep{1.5pt}
    \begin{tabular}{|l|c|}
    \hline
         Maximum Timesteps & 300,000  \\
         \hline
         Learning Rate & 7e-4 \\
         \hline
         Discount Factor & 0.99 \\
         \hline
         Number of Steps & 5 \\
         \hline
         Entropy Coefficient & 0.0 \\
         \hline
         Value Function Coefficient & 0.5 \\
         \hline
         Max. Gradient Norm & 0.5 \\
         \hline
    \end{tabular}
\end{center}
\vspace{-6pt}
\end{table}

\begin{figure}
\vspace{1mm}
\centering
\includegraphics[width=0.9\textwidth]{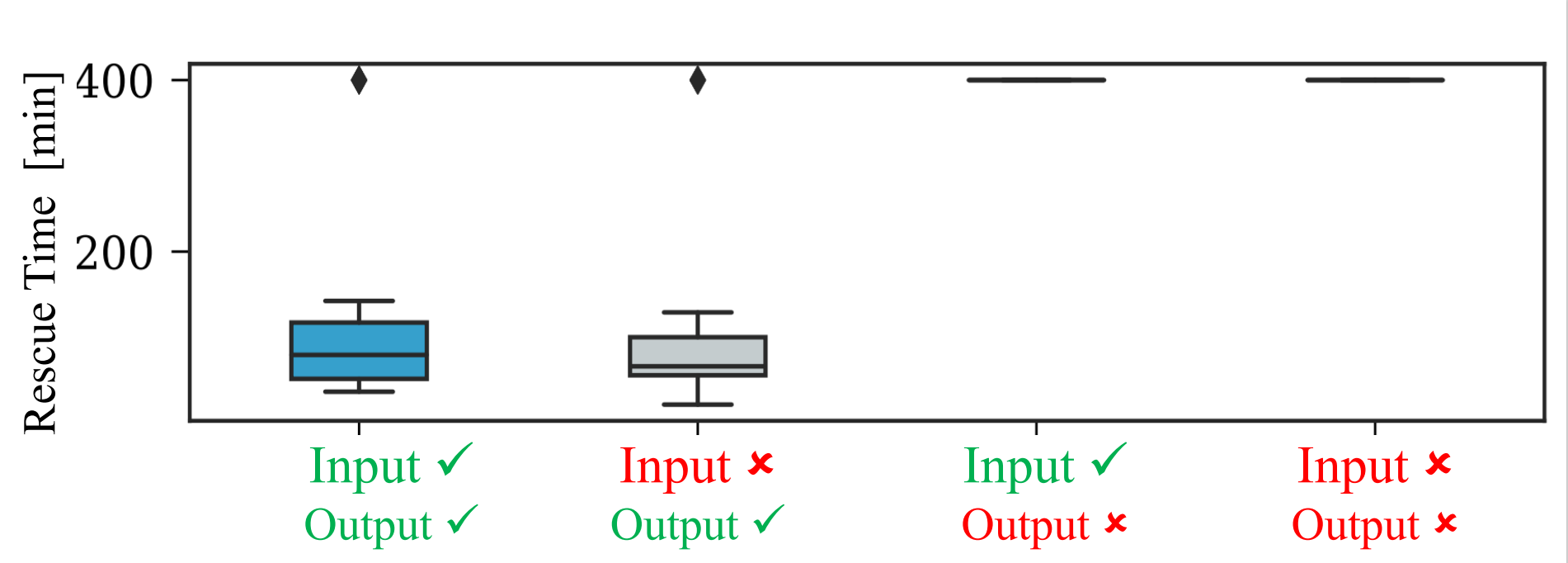}
\caption{The effect of input/output encoding on swarm tactics learning using A2C with 400 minutes time limit for four encoding conditions.} 
\label{fig:RL_rescuetime_long}
\end{figure}

{ Since a much smaller learning cost was allowed to A2C compared to that allowed for Neuroevolution in terms of the number of training samples and computing time (for ease of quick computation), the learned policies of A2C tended to be noticeably poorer. As a result, when testing the A2C generated policies, we increase the maximum allowed mission time to be 400 mins instead of 40 mins. The test results (rescue time) thereof are shown in Fig. \ref{fig:RL_rescuetime_long}. The results obtained via A2C again emphasize the importance of encoding, especially output encoding. Future work will more comprehensively explore the applicability of A2C and other policy gradient methods for learning swarm tactics in such complex environments. }

\section{Conclusion}
\label{sec:Conclusion}
This paper proposes a new computational framework to learn optimal policies for tactical planning of swarm robotic operations in complex urban environments. The tactics can also be learned in scenarios involving uncertain and adversarial conditions. The main contributions of our framework are: 1) neural network-based representation of swarm tactics that encompasses optimal assignment of tasks  and associated swarm primitives to sub-groups; 2) dynamic regrouping of heterogeneous robots for reduced state space representation over a special graph encoding of urban maps; 3) concept of Pareto filtering of points of interest to decrease the state/action dimensionality, which enhances learning tractability. 
We demonstrate the feasibility of learning swarm tactics for search \& rescue missions involving up to 60 UGVs and UAVs, mainly using a neuroevolution algorithm and a preliminary implementation of standard policy gradient RL algorithm (the latter demonstrates the generality of the framework). 
Promising generalizability was observed with levels of performance over unseen test scenarios being comparable to those obtained in training.
This proposed framework can easily translate to other scenarios, swarm settings, and applications. We performed an ablation study highlighting the significance of input/output encoding, which showed that swarm performance decreases substantially when the tactics are learned without encoding state/action spaces. 
                                              

Future work should explore modifying policy gradient methods and combining them with neuroevolution for greater efficiency in tactics generation and subsequent extension to a wider variety of problems involving complex, uncertain environments with additional modalities of perception and action. Furthermore, while the framework readily adapts to varying swarm sizes (demonstrated herein with up to 60 robots), in its current form, it can only deal with a predefined number of squads and Pareto nodes; these restrictions can be relaxed in the future through graph encoding of the tactical action space itself.






\end{document}